\documentclass[aip,jcp,floatfix,preprint,citeautoscript]{revtex4-1}
\setcitestyle{super}

\usepackage{graphicx}
\usepackage{subfigure}
\usepackage{setspace}[2.0]
\usepackage{amsmath, amsthm, amssymb}
\usepackage{xfrac}
\usepackage{dcolumn}
\usepackage{color}

\newcommand{\etal}{\textit{et al.}}

\begin{document}

\date{\today}

\title{Tuning dissociation using isoelectronically doped graphene and
  hexagonal boron nitride: water and other small molecules}

\author{Yasmine S. Al-Hamdani}
\affiliation{Thomas Young Centre and London Centre for Nanotechnology,
  17--19 Gordon Street, London, WC1H 0AH, U.K.}
\affiliation{Department of Chemistry, University College London, 20
  Gordon Street, London, WC1H 0AJ, U.K.}
\author{Dario Alf\`{e}}
\affiliation{Thomas Young Centre and London Centre for Nanotechnology,
  17--19 Gordon Street, London, WC1H 0AH, U.K.}
\affiliation{Department of Earth Sciences, University College London,
  Gower Street, London WC1E 6BT, U.K.}
\author{O. Anatole von Lilienfeld} 
\affiliation{Institute of Physical Chemistry and National Center for
  Computational Design and Discovery of Novel Materials (MARVEL),
  Department of Chemistry, University of Basel, Klingelbergstrasse 80
  CH-4056 Basel, Switzerland}
\author{Angelos Michaelides}
\email{angelos.michaelides@ucl.ac.uk}
\affiliation{Thomas Young Centre and London Centre for Nanotechnology,
  17--19 Gordon Street, London, WC1H 0AH, U.K.}
\affiliation{Department of Physics and Astronomy, 
 University College London, Gower Street, London WC1E 6BT, U.K.}

\begin{abstract}
Novel uses for 2-dimensional materials like graphene and hexagonal
boron nitride (h-BN) are being frequently discovered especially for
membrane and catalysis applications. Still however, a great deal
remains to be understood about the interaction of environmentally and
industrially relevant molecules such as water with these
materials. Taking inspiration from advances in hybridising graphene
and h-BN, we explore using density functional theory, the dissociation
of water, hydrogen, methane, and methanol on graphene, h-BN, and their
isoelectronic doped counterparts: BN doped graphene and C doped h-BN.
We find that doped surfaces are considerably more reactive than their
pristine counterparts and by comparing the reactivity of several small
molecules we develop a general framework for dissociative
adsorption. From this a particularly attractive consequence of
isoelectronic doping emerges: substrates can be doped to enhance their
reactivity specifically towards either polar or non-polar
adsorbates. As such, these substrates are potentially viable
candidates for selective catalysts and membranes, with the implication
that a range of tuneable materials can be designed.
\end{abstract}

\maketitle

\section{Introduction}
Amongst the many materials being studied for chemical applications,
2-dimensional (2D) materials like graphene and hexagonal boron nitride
(h-BN) are some of the most versatile and interesting thanks to their
novel properties and sustainable compositions.  Properties of graphene
and h-BN manifest themselves in various important applications such as
desalination\cite{wang2012water}, water purification\cite{Lei_13},
energy storage\cite{raccichini2015role}, energy
generation\cite{Sales_10,Dhiman_11,Yin_12,Yin_14,Siria_13} and
catalysis\cite{li_2013semi,sun_2011,li2011highly}.  For example,
sizeable voltages have been measured from forming water salinity
gradients across graphene sheets and
nanotubes\cite{Sales_10,Dhiman_11,Yin_12,Yin_14}, and Siria
\etal\ demonstrated that water flowing osmotically through a BN
nanotube produces remarkably large electric currents\cite{Siria_13}.
This was attributed to the possible dissociation and adsorption of
water on the interior of the nanotube which influences the dynamics
inside the nanotube.

Much of the work on graphene and h-BN is also motivated by the
sustainability and the availability of the component elements -- an
aspect which can be difficult to meet using materials that contain
transition or noble metals\cite{fthenakis_sustainability_2009}.
Already, hydrogenated h-BN is thought to be a potential photocatalyst
as a material that is active under visible light and has a band gap
roughly in line with the reduction and oxidation potentials of
water\cite{li_2013semi}. Similar efforts are being made to develop
graphene into a photocatalyst by modification of its band gap, and
also as a support to other photocatalytic
materials\cite{sun_2011,li2011highly}.

Despite the promising applications of h-BN and graphene as membranes
and catalysts, there are still major gaps in our understanding of the
interaction of molecules like water on clean graphene and h-BN
surfaces on the atomic level, and even less is known about how doping
in the materials alters their interaction with molecules.
Indeed, experimental routes to produce hybrid composites of h-BN and
graphene\cite{Ci_2010,Liu_2013} have emerged with high levels of
control being reported on the nanometre scale, which is more reason to
gain better atomic level understanding.  Various theoretical studies
on band gap engineering using h-BN and graphene mixtures
\cite{fan_band_2012,xu_density_2010,shinde_direct_2011,ding_electronic_2009,Ruiqi2012,Nitesh2013,moses2014composition,chang2013band,ferrighi2015boron},
have revealed the tuneability of these materials through the mixture
of atoms. Other studies have focused on exploiting this tuneability
for catalysis of oxygen reduction
reactions\cite{Zhao_2013,baierle_adsorption_2007,kattel_density_2014,li_oxygen_2012,sen_rules_2014,wang_bcn_2012,zhong_nitrogen-_2014,sinthika_doped_2014,wang_vertically_2011,fei_boron-_2014,zhao_can_2013}
and H$_2$
adsorption\cite{baierle_hydrogen_2006,pizzochero2015hydrogen,chhetri2016superior}.

An important aspect to consider, if using graphene and h-BN based
materials as catalysts, is their degree of selectivity.  A high degree
of selectivity is an extremely desirable property for any catalyst and
indeed, the rational design of metal-based heterogeneous catalysts is
the focus of intense research (see \textit{e.g.}  references
\citenum{tsai2015rational,li2015local,yang2013understanding,subbaraman2012trends,nilekar2011mixed,grabow2011mechanism,vojvodic2011optimizing}).
However, even in these cases the metal-based catalysts do not
necessarily have very different selectivities, and although they can
be doped or alloyed to vary their reactivity, the effect on reaction
energies and barriers is often a constant shift with respect to
different
molecules\cite{vojvodic2011optimizing,tsai2014understanding,norskov2002,michaelides2003}.
For instance, in the reaction pathways towards H$_2$ formation
discussed by Cortright \etal, a metal catalyst is used throughout,
which also catalyses H$_2$ consuming reactions
instead\cite{cortright2002hydrogen}. Meanwhile, Guo \etal\ have shown
that a more complex selective catalyst gives rise to a higher
conversion rate of methane to H$_2$
\cite{guo2014direct}.

Here we investigate water and some other environmentally and
industrially relevant small molecules with density functional theory
(DFT). The particular focus of this study is to establish the
thermodynamics of dissociative adsorption and how this is affected by
doping. From this work we draw a number of conclusions. First, doping
strongly affects the dissociation process, in some situations making
dissociation more favourable by several electronvolts. Second,
different surfaces have varying reactivity for the set of molecules
considered, with some substrates significantly enhancing the
reactivity of polar molecules and others enhancing the reactivity of
non-polar adsorbates.



Below, we begin by describing our computational setup in Section
\ref{METHOD} and present our DFT results for water adsorption in
Section \ref{water_result}, followed by an overview regarding the
relative adsorption of other molecules in Section
\ref{molec_result}. In Section \ref{disc} we discuss the trends
observed in adsorption sites and structures, and propose a general
framework for dissociative adsorption before finally concluding, in
Section \ref{CONC}.

\section{Methods}\label{METHOD}
The dissociative adsorption of a water monomer and other molecules on
graphene, h-BN, and their doped counterparts was calculated using DFT
and the Vienna \textit{Ab-Initio} Simulation Package (VASP) 5.3.2
\cite{vasp1,vasp2,vasp3,vasp4}. VASP uses plane-wave basis sets and
projector augmented wave (PAW) potentials\cite{PAW_94,PAW_99} to model
the core region of atoms.

\subsection{System setup}
The graphene and h-BN substrates are modelled using $(5\times5)$
hexagonal unit cells containing 50 atoms, for which adsorption
energies are converged to less than 10 meV with respect to
$(7\times7)$ unit cells.  After a series of convergence tests for the
plane-wave cut-off energy we chose to use a 400 eV energy cut-off,
which gives dissociative adsorption energies converged to within 16
meV of a 600 eV energy cut-off. $\Gamma$-point sampling of reciprocal
space for the (${5\times5}$) cell was used but \textbf{k}-point
densities up to ($9\times9\times1$) were tested. Adsorption energies
using $\Gamma$-point sampling are within 50 meV ($3\%$) of the
converged adsorption energies for all substrates. Spin polarisation
was applied since H pre-adsorption on the substrates gives rise to
spin polarised states. A 10 \AA\ separation in the z-direction between
substrates without a dipole correction proved to be converged for
dissociative adsorption energies of water compared to using a dipole
correction or a 20 \AA\ separation ($<15$ meV
difference).\footnote{For methanol the separation distance in the
  z-direction was increased to 20 \AA\ to allow space for the larger
  adsorbed fragments.}
\begin{figure}
\centering
\includegraphics[width=0.40\textwidth]{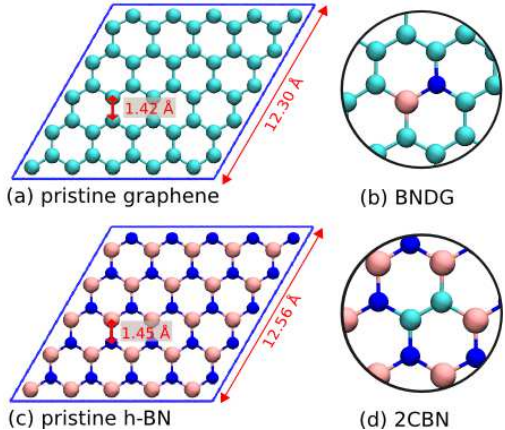}
\caption{The clean and doped graphene and h-BN surfaces considered in
  this study. (a) ($5\times5$) unit cell of graphene. (b) BN doping in
  ($5\times5$) unit cell of graphene which we refer to as BNDG. (c)
  ($5\times5$) unit cell of h-BN. (d) C doping in ($5\times5$) unit
  cell of h-BN, referred to as 2CBN. For clarity only a small portion
  of the ($5\times5$) unit cell is shown in (b) and (d). C is coloured
  cyan, B is pink, and N is blue.}\label{figure_1}
\end{figure}

For the dissociative adsorption energies evaluated here (spanning a
few eV) we have mostly used the Perdew-Burke-Ernzerhof (PBE)\cite{PBE}
generalised gradient approximation exchange-correlation
functional. However we have also verified that the key results
obtained here are not particularly sensitive to the choice of
exchange-correlation functional, as discussed in Section \ref{disc}.

There are many different ways of isoelectronically doping graphene
with BN and \textit{vice versa} and as a first step we focus on low
concentrations of doping: one pair of BN substituting two C atoms in a
$(5\times5)$ unit cell of graphene which we refer to as boron nitride
doped graphene (BNDG) and likewise, two C atoms substituting a BN pair
in a $(5\times5)$ unit cell of h-BN, henceforth referred to as
2CBN. Doped substrates are modelled by isoelectronically doping the
pristine sheets and relaxing the unit cells using a plane-wave energy
cut-off of 600 eV to remove any strain introduced by the mixture of B,
N and C atoms. Relaxation effects are small: less than $1\%$ of the
relaxed lattice constant of the undoped system.\footnote{We verified
  the stability of the doped substrates by calculating their cohesive
  energies and we find good agreement with other work for similar
  arrangements of doping atoms\cite{shinde_direct_2011}. Cohesive
  energies for the different substrates have been calculated as
  $E_{coh}=E^{tot}_{sheet}-{N_C}{E^{tot}_C}-{N_B}{E^{tot}_B}-{N_N}{E^{tot}_N}$
  where $E^{tot}_{sheet}$, $E^{tot}_C$, $E^{tot}_B$ and $E^{tot}_N$
  are the total energies of the sheet and gaseous C, B and N atoms in
  the unit cell, respectively, and $N_C$, $N_B$ and $N_N$ are the
  numbers of C, B and N atoms in the unit cell. The doped sheets in
  this study have cohesive energies between that of graphene and h-BN,
  and the four substrates range between $-7.06$ and $-7.84$ eV/atom.}

When water dissociates on a 2D substrate there are a number of
possible adsorption scenarios. Here, we have focused on four possible
outcomes. Schematic illustrations are given in Fig. \ref{figure_2} and
in brief they involve: (i) An OH group on the surface and the release
of (half) an H$_2$ molecule, referred to as ``OH (\sfrac{1}{2}H$_2$
gas)'';
(ii) The adsorption of both OH and H components of water on the
surface, with them both being on one side of the substrate, namely
``cis(OH--H)''. We consider this configuration to be particularly
important because 2D materials tend to be examined by supporting them
on other materials, leaving only one side of the surface exposed;
(iii) The adsorption of both OH and H on the surface but this time on
opposite sides of the substrate, referred to here as
``trans(OH--H)''. This could arise from having the substrate suspended
in a wet environment or from the H atoms diffusing through the sheet
and there are indications that graphene and h-BN are permeable to
protons\cite{hu2014proton}. However, as it is not clear how likely it
is for molecules to dissociate on different sides of the substrates,
we consider this configuration to be less relevant than cis(OH--H);
(iv) Lastly, ``OH--H--H'' which is again the adsorption of both OH and H,
this time on a surface that has an H atom pre-adsorbed. We tested this
particular set-up in light of previous experimental and simulation
work, where this is thought to cause water
dissociation\cite{Siria_13}.
Many adsorption sites are available for each category and we have
calculated only a number of possibilities: ortho, meta, and para
positioning of the adsorbed components with respect to each other, as
well as adsorption of the components far away from each other and the
doping site in the substrate.

The absolute adsorption energy for dissociative adsorption, $E_{ads}$
is defined as,
\begin{equation}
E_{ads}=E^{tot}_{ads/sub}-E^{tot}_{sub}-E^{tot}_{ads}\label{IE}
\end{equation}
where $E^{tot}_{ads/sub}$ is the total energy of the adsorption
system, $E^{tot}_{sub}$ is the total energy of the relaxed substrate,
and $E^{tot}_{ads}$ is the energy of the intact molecule in the gas
phase. Equation \ref{IE} is used for all but one adsorption state,
that is OH (\sfrac{1}{2}H$_2$ gas). Here we also take into account the
energy ($E^{tot}_{H_2}$) of the \sfrac{1}{2}H$_2$ gas molecule that is
formed:
\begin{equation}
E_{ads}=E^{tot}_{ads/sub}+\sfrac{1}{2}E^{tot}_{H_2}
-E^{tot}_{sub}-E^{tot}_{ads}\label{IE2}
\end{equation}
Within these definitions negative adsorption energies correspond to
favourable (exothermic) adsorption processes.  Bond strengths of
hydrogen and hydroxyl to the surfaces are calculated with respect to a
gas phase hydrogen atom or hydroxyl group instead of the whole
molecule:
\begin{equation}
E_{bond}=E^{tot}_{sub} + E^{tot}_{ads}-E^{tot}_{ads/sub}\label{BS}
\end{equation}

\section{Results}\label{Results}
We begin with the results for the dissociative adsorption of water on
the pure substrates, graphene and h-BN, and on the doped substrates,
BNDG and 2CBN. In general, we find that the dissociation of water is
more facile on the doped substrates and is also strongly affected by
the presence of a pre-adsorbed H atom, local electronic induction, and
steric effects arising from rehybridisation of orbitals in the
substrate atoms. We use these insights to look at the adsorption of
H$_2$, methane, and methanol on the same surfaces in Section
\ref{molec_result}. From our analysis we see that different substrates
favour the dissociation of different molecules, depending on their
polarity, enabling us to make comparisons between the adsorption
behaviour of polar and non-polar molecules and fragments.

\subsection{Dissociative adsorption of water on graphene, h-BN, BNDG and 2CBN}\label{water_result}

\begin{figure}
\centering \includegraphics[width=0.45\textwidth]{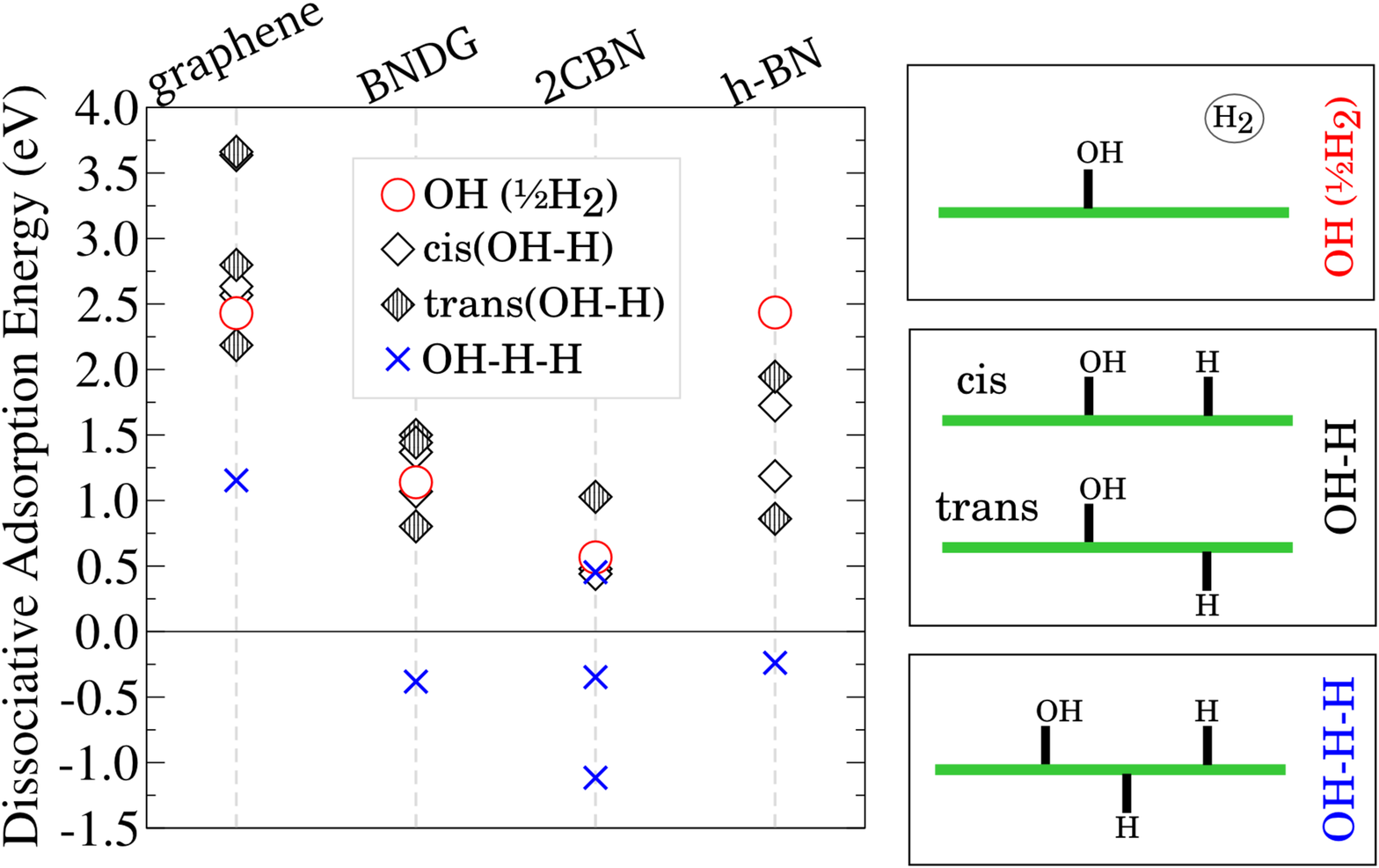}
\caption{The dissociative adsorption energy of water on graphene,
  BNDG, 2CBN and h-BN is shown for different adsorption
  structures. Red circles indicate the adsorption of OH from water
  onto the substrate and the release of hydrogen gas. The black
  diamonds indicate the dissociative adsorption of a water molecule
  into OH and H on the substrate.  The blue crosses correspond to the
  adsorption energies on a hydrogenated substrate. The categories of
  water dissociation on the substrate are illustrated on the
  right.}\label{figure_2}
\end{figure}

Fig. \ref{figure_2} reports results for the dissociative adsorption of
water on the clean and doped substrates. It can be seen that the
energetics of the dissociation process varies significantly for the
various adsorption structures and substrates.

\begin{figure}
\centering \includegraphics[width=0.45\textwidth]{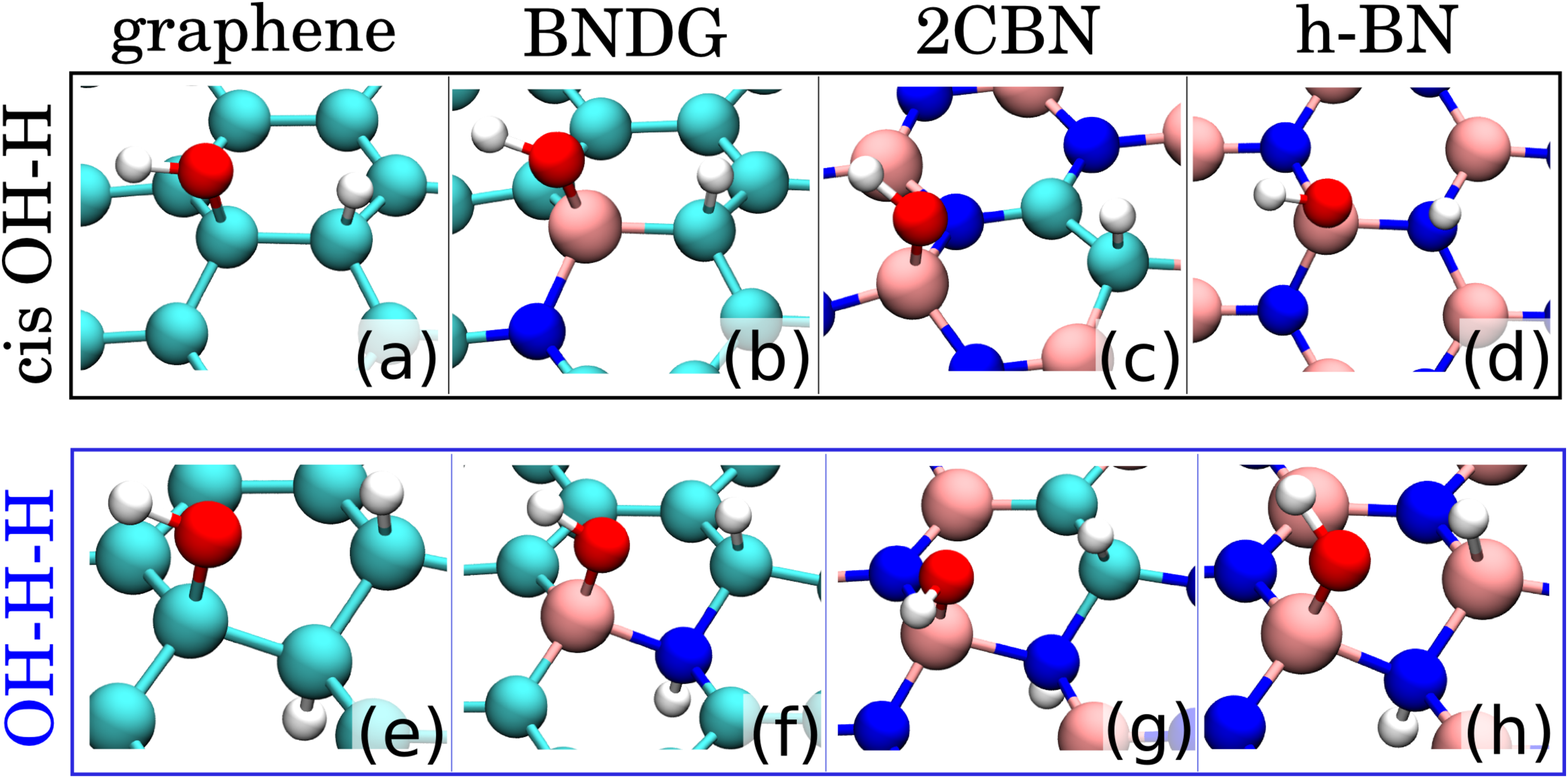}
\caption{The most stable cis(OH--H) (top panel) and most stable
  overall dissociative adsorption structures (lower panel) of water on
  graphene, h-BN, BNDG and 2CBN are shown. (a) and (e) are water on
  pristine and hydrogenated graphene, respectively. (b) and (c) show
  water adsorbed on BNDG and 2CBN, whilst (f) and (g) show water
  adsorption on the hydrogenated counterparts. (d) and (h) are on
  pristine and hydrogenated h-BN, respectively.}\label{diss_fig}
\end{figure}

On pristine graphene we find that dissociation is strongly endothermic
in agreement with previous work\cite{cabrera2007,xu2005water}.  In
addition the energy of the dissociation process varies by as much as 2
eV depending on the final adsorption configuration.  The lowest
adsorption configuration on pristine graphene is trans(OH--H) ($2.19$
eV) with OH and H in ortho positions, in agreement with Xu
\etal\cite{xu2005water} The cis(OH--H) configuration shown in Fig.
\ref{diss_fig} (a) on graphene has a dissociative adsorption energy of
2.57 eV and is thus $\sim0.4$ eV less stable than trans(OH--H).
Dissociative water adsorption is in general, more thermodynamically
favourable on h-BN than on graphene. For example the cis(OH--H) state
on pristine h-BN shown in Fig. \ref{diss_fig} (d), has $E_{ads}$ of
$1.19$ eV and is $1.38$ eV more favourable than the equivalent
structure on graphene. Nonetheless, given just how thermodynamically
unfavourable water dissociation is, it is unlikely that water monomers
will dissociate on pristine graphene and h-BN.

Upon moving to the doped substrates, for which numerous configurations
were considered, we find a significant lowering in the energy to
adsorb water. From graphene to BNDG, and from h-BN to 2CBN, we gain
$\sim1$ eV in the adsorption of a water molecule.  The cis(OH--H)
state and lowest energy dissociation state for each doped surface is
shown in Fig. \ref{diss_fig}. On both BNDG and 2CBN, B--OH and C--H
bonds are formed. Note from Table \ref{bonds} that the B--OH bond is
$\sim1.3$ eV stronger on BNDG than on h-BN (or C--OH on
graphene). Hence, a marked activation of the B atom towards binding OH
results from the mixture of N and C atoms surrounding it and in this
way doping leads to a considerable lowering of the dissociative
adsorption energy.

The presence of the pre-adsorbed H atom also significantly improves
the thermodynamics of water adsorption by $\sim1$ eV for each
substrate.
Favourable OH--H--H configurations are shown in Fig. \ref{diss_fig}
and from Fig. \ref{figure_2} it can be seen that water splitting is
thermodynamically favourable on the hydrogenated h-BN ($-0.24$ eV),
BNDG ($-0.38$ eV) and 2CBN ($-1.12$ eV) surfaces. Thus doping and
hydrogenating both graphene and h-BN makes the thermodynamics of water
dissociation considerably more favourable. The general conclusion that
pre-adsorbed hydrogen facilitates water dissociation is in agreement
with Siria \etal\cite{Siria_13} Interestingly, the overall most
favourable states for water dissociation on the doped surfaces contain
a B--N--C construction in the surface where B--OH, N--H, and C--H
bonds are formed. We considered if the increased reactivity at these
sites is due to the pre-adsorbed H atom on a N site destabilising the
surface and thus activating it towards water adsorption, but this is
unlikely because the N--H bond is very weak (only $0.07$ eV). The
B--N--C construction in the surface of doped substrates is therefore
central to making the dissociation energy more exothermic, and
exemplifies the use of isoelectronic doping to tune the dissociative
adsorption energy of water.
In all OH--H--H states, the OH and H components of the dissociated
water are arranged in a hydrogen bonded fashion. The hydrogen bond on
h-BN at $1.95$ \AA\ is shorter than the hydrogen bond on graphene
($2.23$ \AA) despite the slightly smaller lattice constant of
graphene. The hydrogen bonding distances are indicative of the more
polarised binding of OH and H on h-BN, which culminates in a more
negative oxygen atom in the OH group and hence a shorter hydrogen
bond. 

Additional DFT calculations of water dissociation on the protonated
(as opposed to hydrogenated) substrates were also performed. A
homogeneous background charge is added in the DFT calculations of the
charged systems so that the electrostatic interactions do not diverge
and can be computed under periodic boundary conditions. These reveal
that protonation is slightly less effective than hydrogenation but
still increases the tendency of water to dissociate by $\sim0.8$ eV
with respect to the non-protonated clean surfaces. Thus either
hydrogen pre-adsorption or acidic conditions (pre-adsorbed protons)
could be key elements in the activation of these sheets towards
dissociative water adsorption.

Before moving on to discuss the other adsorbates, two additional
features of these adsorption systems deserve comment. First,
adsorption of the dissociated fragments on separate sides of the sheet
(so-called trans adsorption) is favoured in general. Specifically,
trans-ortho(OH--H) adsorption is $\sim0.4$ eV more stable than
cis-ortho(OH--H) on graphene. This is consistent with previous work on
graphene\cite{lin_hydrogen_2008,boukhvalov2008,balog2010,vsljivanvcanin2011,Merino2015ortho}
and demonstrates the stabilisation gained by adhering to a more
tetrahedral structure around the sp$^3$ hybridised C atom.
Likewise on h-BN and BNDG, the tetrahedral arrangements of
trans(OH--H) and OH--H--H lead to lower dissociative adsorption
energies (by about 0.3 eV). Note the 2CBN system is an exception and
the most stable (OH--H) configuration on 2CBN has cis-para
positioning, shown in Fig. \ref{diss_fig}(c). The trans-ortho (OH--H)
state on 2CBN is still close in energy and only $0.04$ eV less stable
than cis-para\footnote{The 0.04 eV difference between cis-para and
  trans-ortho adsorption configurations remained the same using a
  denser $6 \times 6 \times1$ \textbf{k}-point mesh.}. This can be
explained by the difference in partial charges on the B atoms bonding
to OH in each case. Electronegative N atom neighbours make B atoms
more positive and subsequently form a stronger polar bond with OH. In
the trans-ortho state, the B atom is surrounded by only two N atoms
and hence, is not as electrophilic as the B atom in the cis-para state
which is bonded to three other N atoms. This example in 2CBN
demonstrates that inductive effects from neighbouring atoms dominate
over steric effects. Despite the advantage of satisfying the sp$^3$
hybridisation in trans adsorption states, it is important to remember
that in practice 2D materials are often suspended or grown over
substrates\cite{bn_exp,wang2011monolayer,wang2010periodicity,wintterlin2009graphene,diaz2013hexagonal,altenburg2010graphene,lattelais2015cycloaddition,li2012influence,joshi2012boron}
(metals or silicon carbide) where cis configurations are more likely.

Second, inductive effects are also introduced by the adsorbed water
fragments. This can be seen by comparing the co-adsorbed to the
separately adsorbed OH and H fragments. Specifically, OH
(\sfrac{1}{2}H$_2$ gas) adsorption on graphene and h-BN only differ by
5 meV and indeed the C--OH and B--OH bond (as listed in Table
\ref{bonds}) in graphene and h-BN are almost identical. In contrast
C--H bonds in graphene are significantly stronger than N--H bonds in
h-BN, implying that OH--H on graphene might be more stable, and yet
water adsorption is more exothermic on h-BN. It follows that the
binding of hydrogen atoms on the surface perturbs the local electronic
structure and therefore, the bond strength of OH to the surface, such
that the OH--H configuration is considerably more stable on h-BN than
on graphene.
\begin{table}
\caption{Bond strengths (in eV) for H and OH on graphene, h-BN and
  BNDG sheets with respect to a gas phase hydrogen atom or OH
  molecule. Parentheses indicate neighbouring atoms in the
  substrate. Negative bond energies correspond to endothermic but
  metastable adsorption minima. No minimum was found for OH adsorbed
  on the N atom.}
\begin{tabular}{lr}
\hline\hline
Bond & Bond strength (eV) \\ \hline
Graphene & \\
C--H & 0.81 \\
C--OH & 0.67 \\ \hline
h-BN & \\
N--H & $-0.77$ \\
B--H & $-0.01$ \\
B--OH & 0.67 \\ \hline
BNDG & \\
B--H & 0.98 \\
N--H & 0.07 \\
(B)C--H & 1.15 \\
(N)C--H & 1.04 \\
B--OH & 1.96 \\
(B)C--OH & 0.84 \\
(N)C--OH & 1.03 \\ \hline\hline
\end{tabular}
\label{bonds}
\end{table}

It is useful to explain these trends in terms of the physical
properties of the surfaces and we have done this by looking at Bader
charges\cite{bader1990atoms,henkelman2006fast}, average electrostatic
potentials at each atom, and Kohn-Sham orbitals of the dissociated
states.\footnote{Of course there are many ways to project charges onto
  atoms and Bader charges discussed here are simply used for
  pinpointing the relevant trends in the materials.}  Comparison of
the adsorption structures and Bader charges suggests the most stable
adsorption states arise from: (i) C--H in which the C site has the
most negative partial charge across the surface; (ii) B--OH in which
the B atom is positive and susceptible to nucleophilic attack; and
(iii) N--H in which the N atom is the most negative and therefore the
strongest nucleophile. A careful analysis reveals that the adsorption
of water is affected by a combination of factors involving orbital
overlap and electrostatic interactions. Graphene has weaker
electrostatic interactions with water than h-BN, but better orbital
overlap (evidenced by bond strengths in Table \ref{bonds}). In
contrast, hybrids of h-BN and graphene have stronger electrostatic
interactions with water than graphene, and also stronger orbital
overlap with water than h-BN. Due to these combined effects doped
graphene and h-BN are more suited for the adsorption of
water. Evidently for a given substrate, electrostatic interactions
with a molecule are determining the site of adsorption (\textit{e.g.}
in 2CBN the cis-para state of water is more stable than the
trans-ortho).

To recap, isoelectronic doping has a significant impact on the
thermodynamics of water dissociation of graphene and h-BN. The most
thermodynamically favourable adsorption identified is the OH--H--H
configuration on 2CBN with an adsorption energy of $-1.12$ eV. The
strong adsorption energy on 2CBN can be attributed to: (i) the B--OH
bond in which the B atom is more positive compared to B atoms in the
other substrates; and (ii) a stronger C--H bond at 2CBN as opposed to
a B--H bond at h-BN.

\subsection{Dissociative adsorption of hydrogen, methane, and methanol}\label{molec_result}
\begin{figure}
\centering
\includegraphics[width=0.40\textwidth]{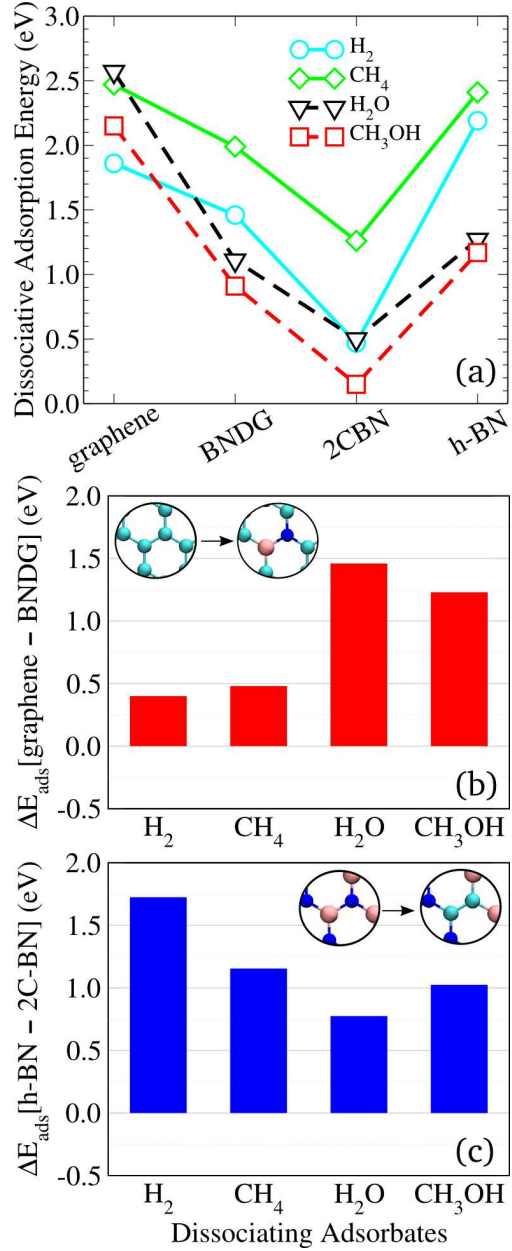}
\caption{(a) Dissociative adsorption energies including ZPE
  contributions of H$_2$, methane, water, and methanol on graphene,
  BNDG, 2CBN and h-BN. H$_2$ in blue circles, methane in green
  diamonds, water in black triangles, and methanol in red
  squares. Results are given only for the most stable adsorption
  structure for each molecule and substrate with the fragments
  adsorbed on the same side of the substrate and without pre-adsorbed
  hydrogen. (b) Gain in adsorption energies from doping pristine
  graphene with BN (in eV) for different molecules, illustrating a
  marked increase in the reactivity with polar adsorbates. (c) Gain in
  adsorption energies from doping pristine h-BN with 2C (in eV) for
  different molecules and here the reactivity with non-polar
  adsorbates increases more significantly. The insets in (b) and (c)
  illustrate the doping.}\label{figure_bar}
\end{figure}

With the insight gained from water adsorption, we also calculated the
dissociative adsorption of H$_2$, methane, and methanol.  As before,
various configurations were calculated for each system and in
Fig. \ref{figure_bar}(a) we report the most favourable dissociation
energies found for the molecules on the same side (cis configurations)
of the pure and doped substrates. The change in zero point energy
(ZPE) upon dissociative adsorption for each system is also included in
the energies in Fig. \ref{figure_bar}. ZPEs were calculated using the
harmonic approximation and we find that the change in ZPE increases
the dissociative adsorption energies by up to $0.3$ eV, which is
certainly not insignificant. In some cases the adsorption energies of
the trans states are more favourable than cis but since it is more
feasible for adsorbates to dissociate on the same side of the
substrate, we show results only for cis configurations. 

From these calculations with the other adsorbates we learn two key
things. First, doping of the pristine substrates helps the
thermodynamics of dissociation for these molecules too. Second, the
details are quite different, with methanol behaving in a similar
manner to water by benefiting most from BN doping in graphene, whereas
H$_2$ and methane benefit most from C doping in h-BN.
Figs. \ref{figure_bar} (b) and (c) illustrate this latter point by
showing the gain in dissociative adsorption energy for each molecule
as a result of doping in the pristine substrates.

The adsorption of methanol varies in a similar way to water across the
different substrates and favours the same adsorption sites (ortho on
graphene, BNDG, and h-BN, and para on 2CBN). From
Fig. \ref{figure_bar} we see that water and methanol adsorption
energies both become more favourable by $\sim 1.4$ eV as the substrate
is changed from graphene to BNDG. Having established that the C--OH to
B--OH change in bond energy is the main contributor to the difference
in adsorption energies for water on graphene and BNDG, we can deduce
that the same is true for methanol. Note that the adsorption of
methanol is stronger than that of water on all substrates by $0.2-0.4$
eV.  On graphene, BNDG, and 2CBN the C--O bond of methanol is broken
preferentially with the CH$_3$ fragment bonding to the substrate at
the same sites as the H from water does. However on h-BN, the O--H
bond is broken instead, resulting in N--H and B--OCH$_3$ bonds with
the h-BN substrate.

Meanwhile the non-polar molecules, H$_2$ and methane, also benefit
from doping of the pristine substrates but in particular from C doping
in h-BN. This appears to be because the alkene-like bond between the
two C atoms, which is susceptible to alkene addition reactions, is
particularly effective at breaking weakly polarised bonds.  Methane
and H$_2$ follow exactly the same trend but H$_2$ is adsorbed around
0.6 eV more strongly overall.

By tracking the lowest adsorption states across the substrates in
Fig. \ref{figure_bar}(a), we see that the preference for H$_2$ and
water switch; H$_2$ adsorbs preferably on graphene and water is
preferred on BNDG and pure h-BN. H$_2$ and water have almost the same
dissociative adsorption energies on 2CBN ($\sim0.5$ eV). The different
adsorption preferences that depend on the isoelectronic substrate
doping is a significant outcome, especially given that these materials
are composed of sustainable and abundant elements, making them
desirable candidates for catalysis.

Finally, as with water adsorption we also examined the effect of H
pre-adsorption on the dissociative adsorption energy of these small
molecules. We found in a similar manner to water that dissociative
adsorption becomes more favourable by $0.7-1.5$ eV on the hydrogenated
surfaces, such that H$_2$, water, and methanol have exothermic
dissociative adsorption energies on BNDG, 2CBN, and h-BN. Thus, as
with water, doping and hydrogenation significantly improves the
energetics of dissociative adsorption on graphene and h-BN.

\section{Discussion and general framework}\label{disc}
Some important trends can be observed from the adsorption structures
and energies of water and the other molecules studied here, which are
likely to apply in general to polar and non-polar adsorbates on BNDG
and C doped h-BN systems. Although we have studied water adsorption
more extensively, the trends also hold for H$_2$, methanol, and
methane. To summarise:
\begin{itemize}
\item Isoelectronic doping of graphene with BN increases the
  reactivity with polar adsorbates (\textit{i.e.} water and methanol)
  by $\sim1.4$ eV but only changes the reactivity with non-polar
  adsorbates by $\sim0.5$ eV. Conversely, isoelectronic doping of h-BN
  with C increases the reactivity most with H$_2$ and methane, by
  $1.2-1.8$ eV.
\item Hydrogen atom (or proton) pre-adsorption on the substrate
  significantly improves the thermodynamics of dissociation for the
  molecules considered by $\sim1$ eV ($\sim0.8$ eV), resulting in
  exothermic dissociative adsorption, and suggesting that acidic
  conditions aid dissociation on the substrates.
\item The most exothermic adsorption sites for polar adsorbates share
  the B--N--C construction, in which there is already a H atom
  pre-adsorbed on a N atom. Meanwhile, non-polar adsorbates favour
  C--C sites with localised electrons (as in 2CBN).
\item Local electronic inductive effects dominate over steric
  effects. In other words, para-positioning of molecule fragments is
  possible (however ortho is generally favoured) if the atoms in the
  substrate have a larger electrostatic potential in the para sites.
\item Atoms in the substrate that change to sp$^3$ hybridisation as a
  result of chemisorption prefer to be in a more tetrahedral
  arrangement, \textit{e.g.} the trans-ortho configuration is
  $\sim0.3$ eV more stable than cis-ortho.
\end{itemize}

Some comments related to these trends are appropriate. First, all the
numbers given have been derived from the PBE exchange-correlation
functional. It is well known that bond strengths and adsorption
energies vary from one functional to the next\cite{schimka2010accurate,vdwpers} and PBE in
particular neglects van der Waals dispersion forces and does not
include exact exchange. Indeed, previous work on similar systems to
those considered here, namely the physisorption of water on
h-BN\cite{al-hamdani2} and on BN doped benzene\cite{al-hamdani1}, has
shown that vdW interactions can be important. Here, however, we are
concerned with strongly bonded chemisorption structures of the
dissociated fragments of water and the other molecules involving an
energy scale of several electronvolts. Nonetheless we have
investigated the dissociative adsorption energies for all states in
Fig. \ref{figure_bar} using the vdW-inclusive optB86b-vdW
functional\cite{vdwDF,B86,vdw_opt11}. We find that the inclusion of
vdW interactions makes the thermodynamics of dissociative adsorption
energy more favourable by $0.2-0.5$ eV. With this functional some
adsorption states are exothermic even in the absence of pre-adsorbed
hydrogen. In contrast, when we look at the thermodynamics of water
adsorption with B3LYP\cite{b3lypA,b3lypB,b3lypC,b3lypD}, that accounts
for some exact exchange but not dispersion, dissociative adsorption is
less favourable by \textit{circa} 0.2 to 0.4 eV. It is clear therefore
that the thermodynamics of dissociative adsorption is sensitive to the
choice of exchange-correlation functional, with the PBE values
presented here resting in the middle of three functionals
considered. Importantly, the relative energies and trends across the
surfaces remains unchanged whether or not dispersion interactions or
exact exchange are accounted for.


Second, when probed experimentally 2D materials like graphene and h-BN
are often adsorbed on a support material such as metals or silicon
carbide. We have not included supporting materials in this study but
the electronic properties of graphene and h-BN can be influenced by
the choice of
support\cite{bn_exp,wang2011monolayer,wang2010periodicity,wintterlin2009graphene,diaz2013hexagonal,altenburg2010graphene,lattelais2015cycloaddition,li2012influence}. Metals
for instance, can hybridise the p$_z$-states in graphene and the N
atoms in h-BN, and thus alter the reactivity of the
surfaces\cite{wintterlin2009graphene,wang2011monolayer,diaz2013hexagonal}. It
is also known that differences in the lattice constants of the 2D
material and support can lead to an undulating moir\'{e} structure in
which different regions of the 2D overlayer interact differently with
the
substrate\cite{bn_exp,wang2010periodicity,altenburg2010graphene,li2012influence,joshi2012boron}. It
would be interesting in future work to explore how the presence of a
substrate alters the trends observed here.

Third, we have seen that depending on the type of doping the
thermodynamics of dissociation of either polar or non-polar molecules
can be enhanced. This would potentially be exploited in heterogeneous
catalysis where it is generally desirable to identify catalysts that
can cleave specific bonds and as a result enhance the selectivity
towards a particular reaction product. In future work it would be
interesting to explore this possibility through calculations of the
kinetics of dissociation on the substrates considered here. However,
since it is now well established that reaction barriers for chemical
reactions at surfaces correlate well with the thermodynamics, it is
not unreasonable to suggest that the thermodynamic trends identified
here could lead to interesting catalytic behaviour.

\section{Conclusion}\label{CONC}
To conclude, the dissociative adsorption of water, H$_2$, methane, and
methanol has been studied on pristine graphene and h-BN, and on their
doped counterparts (BNDG and 2CBN) using DFT. 
Most notably, isoelectronic doping of the pristine surfaces makes the
dissociation process more favourable generally by at least 1 eV. Based
on electronic structure analyses, we conclude that the increased
reactivity of the surface is because B atoms (as a doping species) are
more susceptible to nucleophilic attack, and in 2CBN the C--C double
bond is more susceptible to alkene addition-like reactions. These
changes in the local electronic structure favour particular adsorption
configurations. The OH component bonds strongly to the doping B atom,
whilst H atoms bond preferentially to C compared to either B or N
atoms. Hence, methanol behaves very similarly to water as a polar
molecule, because of the OH group. In the same vein, H$_2$ and methane
follow the same trend across the different surfaces, with both binding
preferentially on 2CBN, where there is a high energy C--C double bond.

The results presented in this study also suggest that adsorption is
exothermic in the presence of adsorbed H atoms (or protons) on the
surface. Thus, there could be important implications for the transport
properties and chemical reactions of water and other molecules across
doped graphene and h-BN membranes, and conditions (acidic or basic)
are likely to be useful gauges for altering the interaction with
molecules. 

Finally, we observe variations in the thermodynamics for the set of
molecules considered depending on the surface. Again we caution that
the calculation of reaction barriers and even rates is an important
next step but these results suggest that one can vary the preference
for H$_2$ dissociative adsorption to that of water or methanol for
example, and consequently alter the course of reaction pathways in
either H$_2$ or methanol formation processes. Consider for example the
wasteful dehydration and methanation reactions in Cortright \etal's
reaction pathways catalysed by a metal for H$_2$
production\cite{cortright2002hydrogen}; wherein H$_2$ is consumed by
reacting with CO$_2$ at low temperatures to produce alkanes and water.
This reaction can be avoided if methanol, methane, and water are split
more readily than H$_2$. According to our findings this might be
achievable for methanol and water by doping graphene with BN.
Overall, our results indicate that isoelectronically doped graphene
and h-BN could exhibit interesting chemical and catalytic activities
which could potentially be exploited.

\acknowledgments We are grateful for support from University College
London and Argonne National Laboratory (ANL) through the Thomas Young
Centre-ANL initiative. Some of the research leading to these results
has received funding from the European Research Council under the
European Union's Seventh Framework Programme (FP/2007-2013) / ERC
Grant Agreement number 616121 (HeteroIce project). A.M. is supported
by the Royal Society through a Wolfson Research Merit
Award. O.A.v.L. acknowledges funding from the Swiss National Science
foundation (No. PP00P2 138932). In addition, we are grateful for
computing resources provided by the London Centre for Nanotechnology
and University College London. We would also like to thank Michail
Stamatakis for his very helpful insights and suggestions.

\end{document}